\documentclass[manuscript,screen]{acmart}
 \pdfoutput=1
\usepackage{dirtytalk}
\usepackage{caption}
\usepackage{subcaption}

\AtBeginDocument{%
  \providecommand\BibTeX{{%
    \normalfont B\kern-0.5em{\scshape i\kern-0.25em b}\kern-0.8em\TeX}}}

\setcopyright{acmcopyright}
\copyrightyear{0000}
\acmYear{0000}
\acmDOI{XXXXXXX.XXXXXXX}

\newcommand\oneA{P01\textsuperscript{Type,High,Ab} }
\newcommand\two{(P02\textsuperscript{Send,Low,Ab}) }
\newcommand\twoA{P02\textsuperscript{Send,Low,Ab} }
\newcommand\three{(P03\textsuperscript{Type,Low,Ex}) }
\newcommand\threeA{P03\textsuperscript{Type,Low,Ex} }

\newcommand\fourA{P04\textsuperscript{Send,High,Ab} }

\newcommand\fiveA{P05\textsuperscript{Type,Low,Ex} }
\newcommand\six{(P06\textsuperscript{Type,High,Ex}) }
\newcommand\sixA{P06\textsuperscript{Type,High,Ex} }
\newcommand\seven{(P07\textsuperscript{Send,Low,Ab}) }

\newcommand\eight{(P08\textsuperscript{Type,Low,Ab}) }
\newcommand\eightA{P08\textsuperscript{Type,Low,Ab} }

\newcommand\nineA{P09\textsuperscript{Send,Low,Ab} }

\newcommand\tenA{P10\textsuperscript{Type,High,Ex} }

\newcommand\elevenA{P11\textsuperscript{Type,Low,Ab} }

\newcommand\twelveA{P12\textsuperscript{Send,Low,Ab} }
\newcommand\thirteen{(P13\textsuperscript{Type,High,Ex}) }
\newcommand\thirteenA{P13\textsuperscript{Type,High,Ex} }

\newcommand\fourteenA{P14\textsuperscript{Type,Low,Ab} }

\newcommand\fifteenA{P15\textsuperscript{Type,Low,Ab} }

\newcommand\sixteenA{P16\textsuperscript{Send,Low,Ex} }

\newcommand\seventeenA{P17\textsuperscript{Send,High,Ex} }

\begin{document}

%%
%% The "title" command has an optional parameter,
%% allowing the author to define a "short title" to be used in page headers.
\title[Key to Kindness: Proactive Content Moderation To Reduce Toxicity Online]{Key to Kindness: Reducing Toxicity In Online Discourse Through Proactive Content Moderation in a Mobile Keyboard}

%%
%% The "author" command and its associated commands are used to define
%% the authors and their affiliations.
%% Of note is the shared affiliation of the first two authors, and the
%% "authornote" and "authornotemark" commands
%% used to denote shared contribution to the research.
\author{Mark Warner}
\email{mark.warner@ucl.ac.uk}
\orcid{0000-0002-7494-6275}
\authornotemark[1]
\affiliation{%
  \institution{University College London}
  \streetaddress{169 Euston Road}
  \city{London}
  \country{United Kingdom}
  \postcode{NW1 2AE}
}

\author{Angelika Strohmayer}
\affiliation{%
  \institution{Northumbria University}
%  \streetaddress{College Street}
  \city{Newcastle Upon Tyne}
  \country{United Kingdom}}
\email{angelika.strohmayer@northumbria.ac.uk}

\author{Matthew Higgs}
\affiliation{%
  \institution{Independent Researcher}
%  \streetaddress{College Street}
%  \city{Newcastle Upon Tyne}
  \country{United Kingdom}}
\email{matthew m@higgs.ac}

\author{Husnain Rafiq}
\affiliation{%
  \institution{Edge Hill University}
%  \streetaddress{College Street}
  \city{Lancashire}
  \country{United Kingdom}}
\email{rafiqh@edgehill.ac.uk}

\author{Liying Yang}
\affiliation{%
  \institution{Northumbria University}
%  \streetaddress{College Street}
  \city{Newcastle Upon Tyne}
  \country{United Kingdom}}
\email{liying2.yang@northumbria.ac.uk}

\author{Lynne Coventry}
\affiliation{%
  \institution{Abertay University}
%  \streetaddress{College Street}
  \city{Dundee}
  \country{United Kingdom}}
\email{l.coventry@abertay.ac.uk}
%%
%% By default, the full list of authors will be used in the page
%% headers. Often, this list is too long, and will overlap
%% other information printed in the page headers. This command allows
%% the author to define a more concise list
%% of authors' names for this purpose.
\renewcommand{\shortauthors}{Warner et al.}

%%
%% The abstract is a short summary of the work to be presented in the
%% article.
\begin{abstract}
Growing evidence shows that proactive content moderation supported by AI can help improve online discourse. However, we know little about designing these systems, how design impacts efficacy and user experience, and how people perceive proactive moderation across public and private platforms. We developed a mobile keyboard with built-in proactive content moderation which we tested ($N=575$) within a semi-functional simulation of a public and private communication platform. Where toxic content was detected, we used different interventions that embedded three design factors: timing, friction, and the presentation of the AI model output. We found moderation to be effective, regardless of the design. However, friction was a source of annoyance while prompts with no friction that occurred during typing were more effective. Follow-up interviews highlight the differences in how these systems are perceived across public and private platforms, and how they can offer more than moderation by acting as educational and communication support tools.

%moderating content prior to publication can
%There is a growing interest in the proactive moderation of content using AI language models to help improve online discourse. Whilst research suggests this moderation paradigm is effective, less is known about how these systems should be designed

% provide context and motivation - what is the problem?

% what did we do to address this?

% what are the main findings

% so what??
\end{abstract}

%%
%% The code below is generated by the tool at http://dl.acm.org/ccs.cfm.
%% Please copy and paste the code instead of the example below.
%%


%%
%% Keywords. The author(s) should pick words that accurately describe
%% the work being presented. Separate the keywords with commas.

\received{20 February 2007}
\received[revised]{12 March 2009}
\received[accepted]{5 June 2009}

%%
%% This command processes the author and affiliation and title
%% information and builds the first part of the formatted document.
\maketitle

\section{Introduction}
% summary background

% key prior work
Artificial intelligence (AI) language models are commonly used to help social networking platforms identify toxic content shared by their users~\cite{gorwa2020algorithmic}. At present, most content moderation systems are reactive as they flag content after it has been shared. However, there is growing interest in applying these models (e.g., perspective API~\cite{perspectiveWeb}) to proactively moderate content to help curb toxic discourse~\cite{katsaros2022reconsidering,chang2022thread,cheng2017anyone}. Proactive moderation could help reduce instances of harmful content being posted and may have positive downstream effects, with fewer toxic messages posted resulting in fewer toxic responses to these posts~\cite{katsaros2022reconsidering,chang2022thread,cheng2017anyone}. Unlike reactive approaches, proactive moderation can operate within private platforms (e.g., chat apps) where it is able to evaluate content on the end user device~\cite{rosenzweig2020law,abelson2021bugs,geierhaas2023attitudes}. 

Proactive moderation approaches typically involve prompting or `nudging' users into making better communication decisions when toxic content is detected. As such, this requires moderation prompts to be designed and embedded within the interaction of the communication system in which it is moderating and so the design of these systems will likely impact how effective they are, and how users interact and perceive them. While emerging research suggests this moderation paradigm is effective~\cite{katsaros2022reconsidering,chang2022thread,schluger2022proactive}, less is known about how the design of these moderation systems may impact efficacy, user communication, and user perceptions. While prior work has explored user perceptions of reactive moderation (e.g.,~\cite{ma2022m,vaccaro2020end}), perceptions of proactive moderation are less well understood. The limited work that has been conducted has focused on a single design approach~\cite{chang2022thread}, and only explored moderation of public communication systems~\cite{katsaros2022reconsidering,chang2022thread,cheng2017anyone}.

%We chose to implement a customised mobile keyboard for text input that temporarily
%disabled the user’s in-built keyboard; a keyboard was chosen as it is platform-independent, and so can be used across
%different communication environments. 

To start addressing these gaps, we developed a semi-functional simulation of a public (Twitter (now 'X', though we continue to use `Twitter' throughout this paper since the research started before the name change)) and private (WhatsApp) communication platform, together with a custom mobile keyboard with embedded proactive moderation. We used this keyboard and our two simulated environments to systematically evaluate three moderation design factors: timing, friction, and the presentation of the AI output. Post-study inquiry-based interviews were conducted to understand why people interacted with the moderation system in the way they did, and what their perceptions are of this approach to moderation across both public and private platforms, and across different designs. This work contributes new insights into the effectiveness of proactive moderation design factors through a controlled online experiment. Moreover, through our follow-up interviews, we provide new insights into proactive moderation perceptions, building on prior work by exploring design practices and perceptions across both public and private communication contexts. The key findings from our work are as follows: 

\begin{enumerate}
  \item[-] Proactive moderation was effective across public and private platforms, regardless of the design.
  \item[-] Friction was a source of annoyance, and where friction was low \textit{on typing} timing prompts were more effective.
  \item[-] People used `phantom messages' (i.e., messages that are never sent) to express their anger or frustration. \textit{During typing} interventions attracted heightened privacy concerns due to `phantom messages' being moderated. 
  \item[-] Proactive moderation can help to support people in their communications and act as an educational tool which has the potential to result in longer-term behaviour change.
  \item[-] Concerns exist over the accuracy, usability and privacy of the system, and how it may impact communications; with concerns more prominent within private platforms. 
\end{enumerate}

\begin{figure}[h!]
  \includegraphics[width=\textwidth]{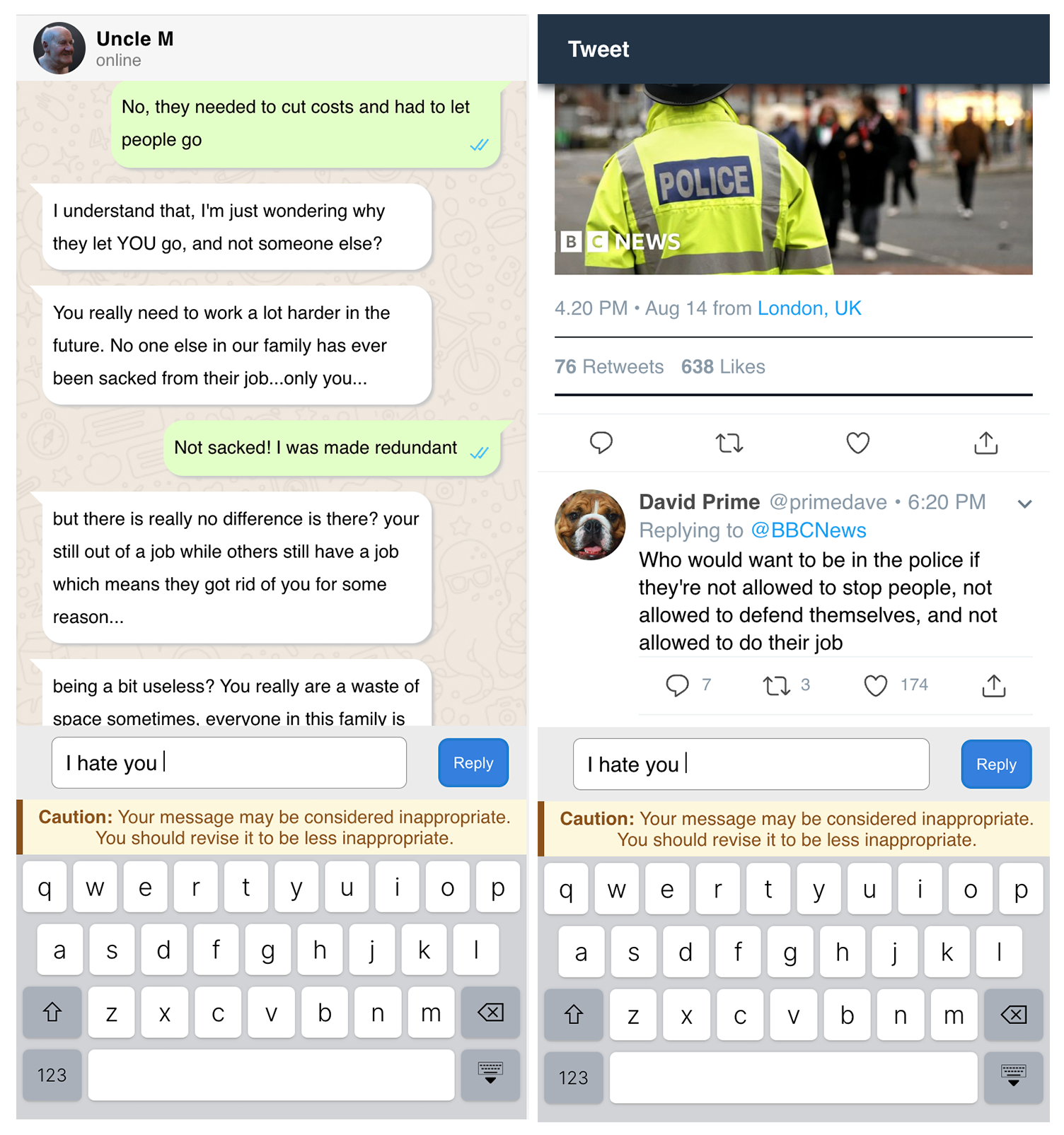}
  \caption{Example of the two simulated platforms, with WhatsApp shown on the left, and Twitter shown on the right. The moderation prompt in these two examples is triggered on typing, includes no friction, and presents an abstract message. The message that triggered the prompt was 'I hate you' which we use for illustrative purposes only, to avoid reproducing overly toxic or hateful content.}
  \Description{}
  \label{fig:examples}
\end{figure}

\section{Background}

\subsection{Content moderation complexities in private vs. public platforms}
There are some important differences between private and public platforms that impact content moderation. Public platforms like Twitter are social networking platforms that promote the presence of communities through more public sharing of content, while private platforms like WhatsApp are chat apps that promote secure and private (i.e., end-to-end encrypted) messaging between individuals and groups. One distinction between these two types of platforms is in where the communications are stored, with social networking platforms storing data on servers and chat apps storing data on end-user devices~\cite{abelson2021bugs,rosenzweig2020law,geierhaas2023attitudes}. Toxic content is a problem on both public (e.g.,~\cite{chatzakou2017hate, mittos2020a}) and private platforms (e.g.,~\cite{sanchez2022whatsapp,sultana2021unmochon}), and Scheuerman et al.~\cite{scheuerman2021framework} found that some harms that occur in private ``spheres'' result in those being targeted feeling more isolated and alone, potentially exasperating harms. 

Despite the harms known to occur on private platforms, far more content moderation occurs on public platforms as the content is accessible to platform owners. While public platforms all take different approaches to how they moderate their online environments though differences in their community guidelines~\cite{jiang2020characterizing}, most recognise the need for some level of moderation to prevent harm~\cite{gillespie2018custodians}, even fringe platforms such as Gab~\cite{blekkenhorst2019moderating}. Moderation on these platforms typically involves human and AI review, with platforms applying either hard or soft moderation approaches. Hard moderation~\cite{zannettou2021won} often involves removing content that is deemed to violate community guidelines. Most platforms also have account restriction policies that include account suspensions, deplatforming~\cite{ali2021understanding,jhaver2021evaluating}, and restricting account and content visibility (i.e., shadowbanning)~\cite{are2022shadowban,gillespie2022not,are2023autoethnography}. Soft moderation is also applied within the reactive moderation paradigm, for example by applying context labels to posts~\cite{allen2022birds,jones2022misleading}.

These reactive moderation approaches do not reduce the harm caused by content before its removal. Moreover, as the approach relies heavily on human review, there are significant hidden labour costs~\cite{steiger2021,roberts2014behind,roberts2016commercial,gillespie2018custodians,dosono2019moderation}. Reactive moderation and its associated response policies may also disproportionately affect certain groups such as transgender users~\cite{haimson2021disproportionate,dias2021fighting}, black users~\cite{haimson2021disproportionate,marshall2021algorithmic}, and users posting nudity and sexuality~\cite{are2023autoethnography}. There are also concerns around the lack of transparency of moderation systems~\cite{jhaver2019does,gorwa2020algorithmic}, and the lack of usable recourse mechanisms available to those experiencing account and content restrictions~\cite{are2023autoethnography}. Kiritchenko et al.~\cite{kiritchenko2021confronting} argue that using similar language models to proactively moderate content could help alleviate many of these concerns~\cite{kiritchenko2021confronting}. Moreover, with emerging technologies that scan content stored on end-user devices, an approach known as client-side scanning (CSS)~\cite{abelson2021bugs,rosenzweig2020law,geierhaas2023attitudes}, these previously un-moderated spaces could become subject to moderation. %CSS is a novel approach as it arguably circumvents the encryption built into private communication platforms by scanning the content in an unencrypted state on the end user device, allowing for the moderation of content within private communication platforms. 
CSS has primarily been explored around child sexual abuse images (CSAM)~\cite{REPHRAIN}, and while it provides a potential solution to the moderation of private platforms, there are concerns over its potential to be misused as a surveillance tool, and encroach on user privacy~\cite{abelson2021bugs,rosenzweig2020law}. One study conducted with German citizens found that people were only willing to accept this form of moderation to detect serious and organised criminal activity (e.g., CSAM, terrorism)~\cite{geierhaas2023attitudes}. Yet, most concerns relate to how these systems propose to report detected content to third parties (e.g., platforms, police). CSS to detect and moderate toxic content does not involve third-party reporting, relying instead on \textit{soft} moderation approaches that `nudge' users~\cite{thaler2009nudge} towards reducing toxicity within their communications.

\subsection{Proactive content moderation}
Proactive content moderation is typically a \textit{soft} moderation approach~\cite{katsaros2022reconsidering} that does not involve the removal of content or the restriction of accounts but instead raises awareness around problematic content to encourage people to communicate in a more pro-social way. It does this by attaching informative warning labels to draft messages before sending~\cite{seering2019moderator, chang2022thread,van2017thinking,jones2012reflective,vincent2020twitter, katsaros2022reconsidering,wang2013privacy}. These approaches often~\cite{schluger2022proactive} rely on AI language models to detect problematic content, whilst having the potential to alleviate many of the issues and concerns present when these models are used within the reactive moderation paradigm~\cite{kiritchenko2021confronting,gorwa2020algorithmic}.

Some of the earliest strands of work exploring content moderation that could be considered proactive were those interested in supporting users in privacy and disclosure decisions online. Wang et al.~\cite{wang2013privacy,wang2014field} explored the application of nudge theory~\cite{thaler2009nudge} to enhance users' awareness of their privacy. They developed three nudges, one to raise awareness of the user audience, one to slow down the interaction to reduce ``heat of the moment'' posts, and one to raise awareness of the sentiment within posts. Their qualitative evaluation of these nudges found the first two to be perceived as useful and to have a positive impact on how users evaluated messages prior to posting. However, delaying users' posts received mixed results with some users perceiving the benefits, and others becoming annoyed. Similar work has explored explicit interface cues that indicate community disclosure norms, finding that users respond to these cues by mimicking disclosure norms learnt through the cues~\cite{spottswood2017should}. In more recent work, Schluger et al.~\cite{schluger2022proactive} discuss two different forms of proactive moderation: static and dynamic. Static includes design choices that promote pro-social behaviour that does not involve any direct interaction between moderators and users. Examples include the listing of community rules or guidelines to help users learn the norms of a community~\cite{dym2020social}; promoting pro-social behaviour by embedding signalling mechanisms into the design of platforms as was seen in dating apps used by gay and bisexual men, to promote ``living stigma free''~\cite{warner2018privacy,levy2017designing}; and priming users towards more pro-social interactions through changes to the design of the user interface~\cite{seering2019designing}. Dynamic approaches involve direct engagement by moderators, but with advances in AI language models, there is a growing trend towards dynamic proactive moderation systems being automated. This has the advantage of reducing the human labour costs associated with human moderation and reducing users' exposure to problematic content. 

We have started to see this moderation approach being explored in academic research~\cite{seering2019moderator, chang2022thread,van2017thinking,jones2012reflective} and used on platforms~\cite{vincent2020twitter, katsaros2022reconsidering}. Schluger et al. designed an algorithmic tool to support moderators in proactively moderating Wikipedia Talk Page. They found moderators already engaging in proactive moderation behaviours, and that the proactive intervention tool resulted in more nuances (or `soft') moderation actions, as opposed to hard moderation actions used when employing reactive moderation strategies. Chang et al. ~\cite{chang2022thread} developed a browser plug-in that provided users with conversational guidance. On selecting `reply' when conversing in the r/ChangeMyView subreddit, the plug-in indicated the risk of a conversation turning uncivil, and how their draft reply may impact the conversation's incivility risk. Similarly to the findings from Schluger et al., they found participants (in this case users, not moderators) were already performing the tasks of the intervention - which in this case was the evaluation of conversation incivility risk - yet the intervention tool was still seen as valuable. The experiment suggests that the intervention was effective when it warned users that the conversation was at risk of becoming uncivil. We have also seen proactive moderation interventions embedded into live social networking systems, such as Twitter. An evaluation of this intervention found it to be effective at reducing offensive Tweets, with 9\% being deleted and 22\% being revised. The results also suggest possible downstream effects, with fewer toxic messages present on the platform resulting in fewer toxic messages that would have otherwise been posted in reply or in response to those deleted or revised messages~\cite{katsaros2022reconsidering}; an effect also suggested by Chang et al.~\cite{chang2022thread}.

\subsection{Design of proactive moderation systems}\label{background:design}
As our research aims to systematically investigate the design of proactive moderation systems, this section explores research into proactive moderation and warning mechanisms within communication systems, to identify salient design elements from prior interventions. Prior work suggests that people often make poor communication decisions when they are experiencing negative and heightened emotions~\cite{wang2011regretted,sleeper2013,warner2021oops}. Moreover, Cheng et al.~\cite{cheng2017anyone} reported that a primary trigger for trolling behaviour can be a negative mood ~\cite{cheng2017anyone}. Acquisti~\cite{acquisti2004privacy} highlights how people often find it difficult to engage in self-control online due to the pull of immediate gratification through the sharing of online content, even when they may know the risks involved in sharing. In Wang et al.'s research~\cite{wang2013privacy,wang2014field}, they explore the use of a time delay nudge to encourage reflection around content being drafted, to help reduce inappropriate disclosures. Masrani et al.~\cite{masrani2023slowing} explored the use of friction to reduce polarisation in online discourse by limiting participants to sending one message per two-minute interval using Discord's ``slow mode" feature. Similarly to~\cite{wang2013privacy,wang2014field}, they found it to frustrate participants; yet they also found it to result in participants writing more thoughtful messages. Prior work by [hidden for review] which explored proactive content moderation within design workshops with a variety of stakeholder groups, identified friction and resistance as a potential way to encourage user reflection. These design mechanisms are intended to act as design frictions that can help ``disrupt `mindless' automatic interactions, prompting moments of reflection and more `mindful' interaction''~\cite{cox2016}. 

The frictions that have been previously explored all occur at a specific point during the interaction (i.e., directly before a message is sent). Yet, prior research exploring content-based prompts shows that the timing of when prompts are activated can impact their effectiveness. Prior work has shown this around privacy indicators~\cite{egelman2009timing,shulman2022informing,balebako2015impact}, prompts within self-regulated learning environments~\cite{thillmann2009merely}, and in digital health interventions~\cite{alkhaldi2016effectiveness}. Moreover, in design research conducted by [hidden for review], they found participants discussing proactive moderation interventions at different points during the intervention, both during typing and before sending. No prior research has explored the relationship between the timing of moderation prompts, and the presence of friction. As prior work~\cite{wang2013privacy,wang2014field,masrani2023slowing} shows how users can become frustrated with frictions, it would be important to understand whether shifts in the timing of a prompt that incorporated friction may limit these feelings of frustration. 

%The last design element we explore is the element of presentation. 
In exploring the broader literature on moderation and warning prompts, we identified factors related to how the output of AI models are presented to users. While it is undoubtedly important to understand the characteristics of the data used to train algorithms and how algorithms transform data to make predictions, it is also important to understand how the outputs of algorithmic systems are presented to users, and how different presentations impact user interpretation and behaviour~\cite{van2023measurements}. The limited research that has explored model presentation within the proactive moderation context shows how these systems could foster anti-social rather than pro-social behaviour, through misuse of the system due to model presentation. A particular concern was raised related to how an algorithm that presented its output in detail could lead to gamification of abuse (e.g., users trying to get a high toxic score) [hidden for review]. Research findings on the presentation of warning prompts more broadly suggest that presentation should be considered in relation to the goals of the prompt. For example, within the context of phishing warnings which require quick evaluations of risk, Lain et al.~\cite{lain2022phishing} found that more detailed warnings were no more effective than simple ones. In contrast, within the context of misinformation labels which are designed to counter false information and avoid harmful belief formation, detailed messages were found to be more effective~\cite{ecker2010explicit,swire2017processing,chan2017debunking}.

%\subsection{Designing content moderation warnings/prompts}

\subsection{Summary and research questions}
Prior work provides useful insights into proactive content moderation interventions that occur before a user publishes content. Most of the interventions show improvements in awareness and user behaviour when the intervention is applied. We see interventions like this being designed in different ways, but most include an algorithmic evaluation of the message, and awareness mechanisms, whilst others include design frictions~\cite{cox2016} intended to slow down the interaction. Most of the intervention mechanisms focus on interactions within public social media platforms like Facebook~\cite{wang2013facebook,wang2014field} and Twitter~\cite{katsaros2022reconsidering} with no known prior research exploring the application of these mechanisms in private chat apps. Moreover, prior research has not systematically investigated the design of proactive moderation systems in such a way that considers how design factors interact, and in particular how timing, friction, and presentation impact the effectiveness and user experience around proactive moderation. As such, we pose the following research questions that help to guide our research towards addressing some of these open questions:

\begin{enumerate}
  \item[\textbf{RQ1}] How effective are proactive moderation prompts at reducing message toxicity across public and private platforms?
  \item[\textbf{RQ2}] How do timing, friction, and model presentation affect intervention effectiveness and how do these factors interact? 
  \item[\textbf{RQ3}] How do people perceive proactive moderation systems when implemented at the point of us (e.g. within a keyboard), and how does it impact communications?
\end{enumerate}

\begin{figure}[p]
  \includegraphics[width=\textwidth]{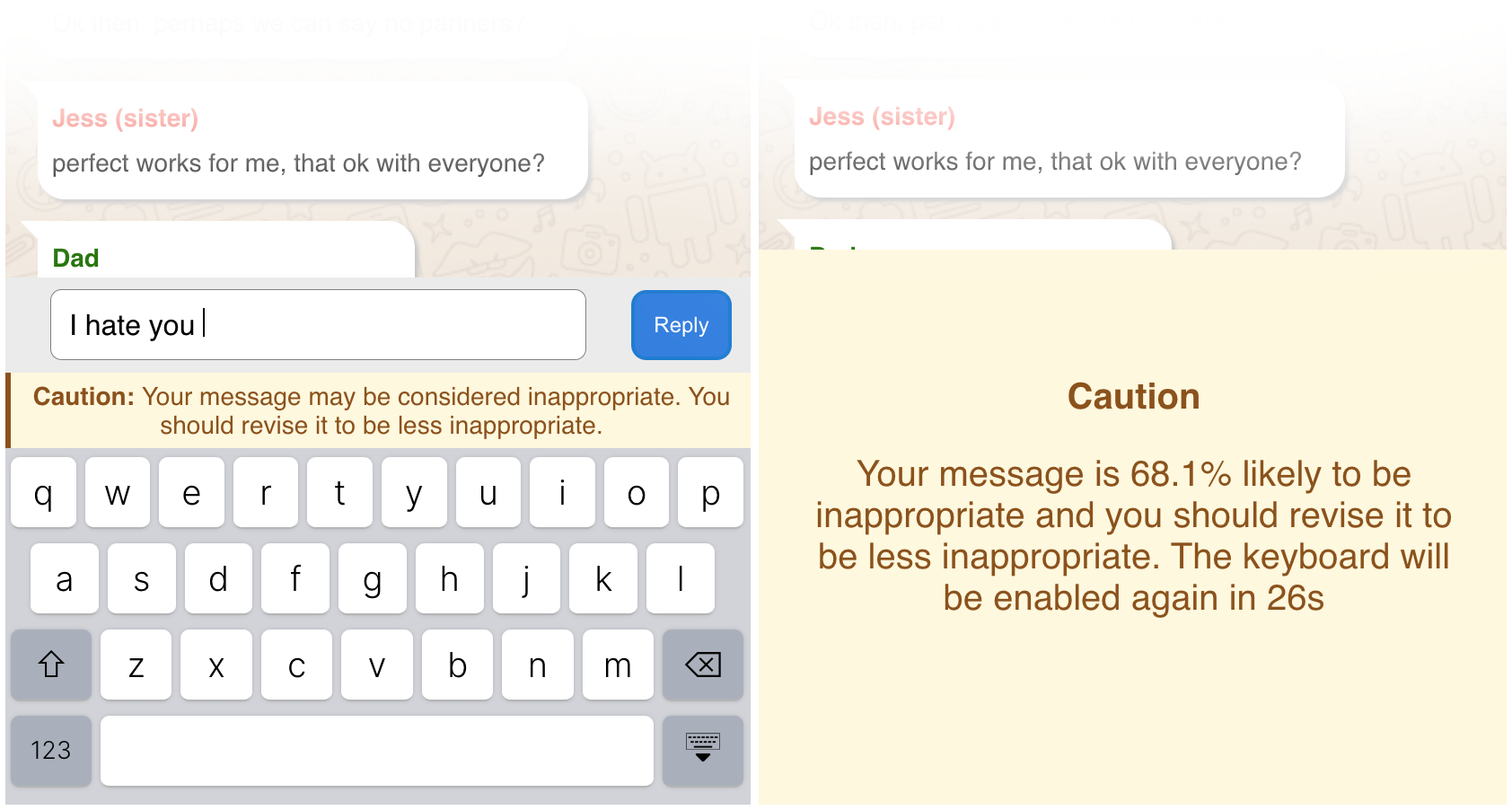}
  \caption{Example of two~\textit{on typing} keyboard moderation prompts that have occurred during a WhatsApp conversation. The left shows a prompt with low friction with an abstract presentation of the model. The right shows a prompt with high friction (keyboard disabled for 30 seconds) and an explicit presentation of the model. The message that triggered the prompt was 'I hate you' which we use for illustrative purposes only, to avoid reproducing overly toxic or hateful content.}
  \Description{}
  \label{fig:onTyping}
  \includegraphics[width=\textwidth]{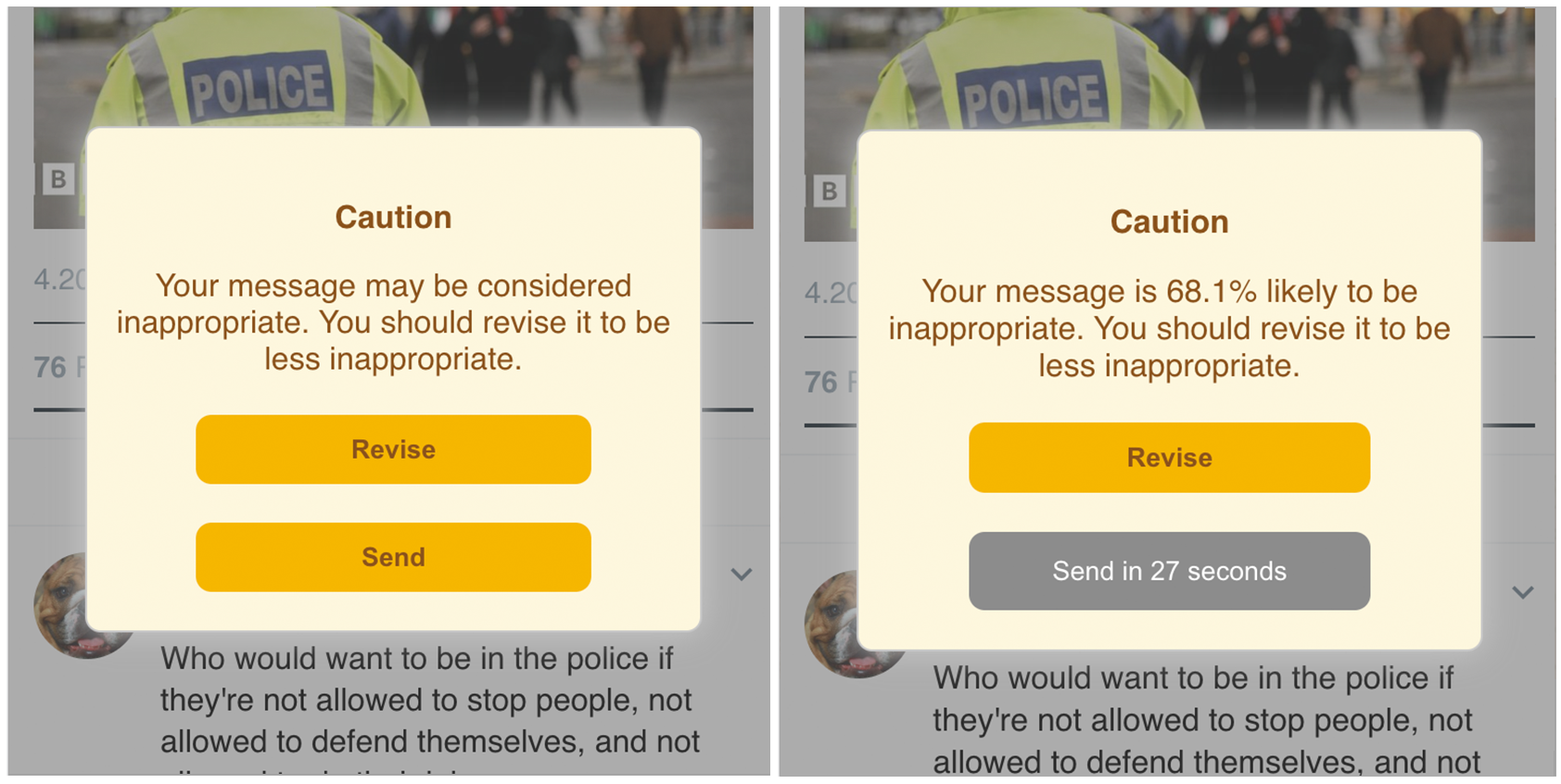}
  \caption{Example of two~\textit{on send} keyboard moderation prompts that have occurred in a Twitter thread response. The left shows a prompt with low friction with an abstract presentation of the model. The right shows a prompt with high friction (send button disabled for 30 seconds) and an explicit presentation of the model.}
  \Description{}
  \label{fig:onSend}
\end{figure}

\section{Experimental platform design}
We developed a semi-functional simulation of a public (Twitter) and private (WhatsApp) communication platform within our experimental environment (see Fig~\ref{fig:examples}). On top of this, we implemented a customised mobile keyboard with built-in proactive moderation. This keyboard was used for text input, and temporarily disabled the user's in-built keyboard; a keyboard was chosen as it is platform-independent, and so can be used across different communication environments. This keyboard and the simulated environments were used to systematically evaluate proactive moderation design factors.

%This allowed us to build different proactive moderation intervention features into the keyboard, and evaluate them within a controlled setting. 
%Below, we detail the design factors that played into our system (timing, friction, and model presentation). Furthermore, we explain our process of developing the stimuli that we asked participants to respond to during the experiment. 
To evaluate the toxicity of messages being typed and to determine when a moderation event should occur, we used Perspective API~\cite{perspectiveWeb} with a threshold of 60\%. We chose this threshold as it captured more borderline cases of toxic speech, and allowed us to understand how false-positive or less obvious forms of toxicity were experienced by users within this moderation paradigm. To develop our keyboard intervention, we used insights gained from prior literature explored in section~\ref{background:design}. Below we detail how we implemented our experimental conditions.

\subsection{Prototype design}
After a review of the literature, we identified three design factors that may influence the effectiveness and use of proactive moderation systems across platforms. In total, we implemented eight experimental conditions (timing(2) vs friction(2) vs presentation(2)), and a control condition where no moderation was active.

\subsubsection*{Timing}Two timing conditions were identified, the first being~\textit{on typing} which prompted participants if toxicity was detected in their draft message during typing (see: Fig~\ref{fig:onTyping}) and the second being~\textit{on send} which prompted participants after pressing `send' (see: Fig~\ref{fig:onSend}).

\subsubsection*{Friction}In applying friction, we aimed to understand how this may help create reflection and encourage less toxicity when typing, and how friction interacts with the other design factors. Within our~\textit{high friction} designs (see: Fig~\ref{fig:onTyping}, right)), we used a 30-second timer to restrict functionality, while in the~\textit{low friction} designs, no functionality was restricted (see: Fig~\ref{fig:onTyping}, left). To ensure the 30-second delay was meaningful, we conducted several rounds of in-person and remote pilots where we adjusted the timer from 10 to 60 seconds, observing interaction and collecting qualitative feedback from participants. Using these pilots we qualitatively evaluated the friction, determining that 30 seconds was not overly long to be overly frustrating, yet long enough to be noticeable.

\subsubsection*{Presentation}We implemented a short descriptive warning within our prompt design, in line with prior work~\cite{lain2022phishing} that shows how short prompts are more effective where the goal of the user is to quickly evaluate risk. To understand the impact of model~\textit{presentation} within the prompt, we either~\textit{abstracted} the model's output by presenting a generic prompt (see: Fig~\ref{fig:onTyping}, left), or made the prompt~\textit{explicit} by presenting the \% likelihood score (see Fig~\ref{fig:onTyping}, right).

\subsubsection*{Timing and friction interaction}
Where a moderation event occurred in the \textit{on send} timing condition, a pop-up dialogue appeared informing participants that their message ``may be considered inappropriate", and asked them to revise their message. Where friction was high, the `send' button was disabled for 30 seconds, where friction was low the send button remained enabled. Where a moderation event occurred in the \textit{on typing} timing condition, where friction was high the keyboard would disable for 30 seconds with the user being informed that their message ``may be considered inappropriate" (see Fig~\ref{fig:onTyping} (right)). In the low friction condition, a non-intrusive information banner would appear at the top of the keyboard, showing the same message as above (see Fig~\ref{fig:onTyping} (left)).

\subsubsection*{Triggering of moderation events}
In the \textit{on typing} timing condition, messages were evaluated by the API after each typed word was entered into the keyboard. If the model evaluated the draft message as toxic (i.e., the API returned a value above the threshold), it would trigger a moderation event. No new events would occur unless the message switched from non-toxic, and then back to being toxic. In the \textit{on send} timing condition, draft messages were evaluated once the participant pressed `send'. If they chose to revise their message, the message would move back into a draft state to allow the participant to make revisions. It would be evaluated again by the API once the participant pressed `send'. This would continue until the participant sent their message. 

\subsection{Message and thread stimuli development}
Participants were asked to respond to six message threads (three simulated on WhatsApp, and three on Twitter). These were developed to provoke an emotional response from participants and to increase the likelihood of moderation exposure. To develop the Twitter stimuli, we analysed real Tweets, evaluating the toxicity of replies to Tweets. As we did not have access to WhatsApp messaging data, we collaboratively developed our own set of stimuli which we evaluated and iterated on through a series of pre-studies. This section details the methods we developed to create these stimuli. 

\subsubsection{Twitter conversations}
To create our Twitter stimuli, we selected the Twitter handle of a news organisation (@BBCNews) that aims to be impartial and regularly posts news articles covering a wide variety of topics. Using the Twitter API~\cite{TwitterAPI}, we downloaded the timeline of @BBCNews and identified conversations likely to provoke a toxic response from our participants. For each Twitter stimuli, we included a ``seed'' tweet (the original @BBCNews post) and a provocative reply, as we found the most provocative replies to be more provocative than the seed tweet alone. We considered only the top 30 potential stimuli in the timeline of @BBCNews which we ranked according to the number of subsequent replies. To estimate the probability that potential stimuli would provoke a toxic response from a participant, we used the Perspective API~\cite{perspectiveWeb} to count the number of toxic replies to each stimuli and used Bayesian inference to estimate the probability that a future reply would be toxic. We then collaboratively identified a final set of stimuli containing a high probability of provoking a toxic response \emph{and} with seed tweets (i.e. @BBCNews posts) covering a diverse range of topics.

\subsubsection{WhatsApp message threads}
To create our initial WhatsApp stimuli, we collaboratively developed a set of relationship types (e.g., family, work) and conversational dimensions that may provoke an emotional response (e.g., family dispute, professional understating). From these, we further collaborated to develop an initial set of 10 conversation threads and corresponding scenarios which we evaluated in three online pre-studies (pre-study 1 ($N=50$), pre-study 2 ($N=51$), pre-study 3 ($N=30$)). Each pre-study round involved a gender-balanced pool of participants. Between each pre-study, we iterated on the threads based on feedback.  We collected information on how likely participants were to respond to the message threads as we wanted threads that would provoke a response. We also asked how likely they would be to respond with a toxic message by asking ``How likely would you be to respond to this message in an angry or aggressive way (e.g., use of aggressive or foul language)?''. Finally, we asked them for feedback on how we could increase the likelihood of them responding in a more toxic manner. After three rounds of refinement, we selected the four threads with the highest combined scores of (1) likelihood to respond and (2) likelihood to respond in a toxic way. Our final stimuli set contained a pairwise chat in a family relationship context, a pairwise chat in an intimate relationship context, and a group chat based in a family relationship context. 

\section{Method}
The work discussed in this paper took place across two linked studies. First, we conducted an online experiment to quantitatively test people's use of and responses to our keyboard designs. Following this, we invited people across all design groups to participate in an interview. This allowed us to provide answers to our research questions in a way that indicates generalisable information on the impact of certain design factors as well as more in-depth learning about their impacts on people more broadly beyond the text they were typing. We used a semi-functional simulated environment as opposed to an ``in the wild'' deployment due to technical constraints with running a suitable ML model locally on a mobile device, and the privacy and data ethics concerns with using an external API to compute toxicity scores. While it would be important to understand real-world usage, especially over a longer period, this study aimed to develop an initial understanding of usage before further development. 

\subsection{Study 1 (Online experiment)}
The online experiment aimed to understand the effectiveness of different proactive moderation designs that incorporated our three design factors: friction, timing, and model presentation. We evaluated effectiveness by analysing the change in message toxicity score, and the likelihood that a message would be revised. 

\subsubsection{Participant recruitment}
Between 13 - 17 December 2022, we conducted an online study that involved participants ($N=575$) over the age of 18 living in the UK, being asked to respond to a series of Tweets and WhatsApp messages using a custom keyboard (see Fig~\ref{fig:onTyping} and~\ref{fig:onSend}). Participants were recruited via Prolific.ac, with gender balancing. An overview of our sample demographics can be found in Table~\ref{tab:demographics_online}. Participants were randomly assigned to an experimental group, with a control group being used as a baseline.

\begin{table}[]
\centering
    \caption{Demographics of our online experiment sample ($N=575$) and the sub-sample ($N=165$) of those that triggered at least one moderation event during the study.}
    \label{tab:demographics_online}
\begin{tabular}{lll}
\toprule
\textbf{Gender}                             & All & Event \\
\midrule
\hspace{0.2em} Male                                        & 283 & 100        \\
\hspace{0.2em} Female                                      & 284 & 61       \\
\hspace{0.2em} Non-binary                                  & 4   & 1         \\
\hspace{0.2em} Undisclosed                                 & 4   & 3         \\
\textbf{Age}                                &     &           \\
\midrule
\hspace{0.2em} Under 24 years old                          & 61  & 12        \\
\hspace{0.2em} 24 - 34 years old                           & 190 & 62        \\
\hspace{0.2em} 35 - 44 years old                           & 175 & 55        \\
\hspace{0.2em} 45 - 54 years old                           & 92  & 24        \\
\hspace{0.2em} 55 - 64 years old                           & 44  & 8         \\
\hspace{0.2em} 65 years or over                            & 9   & 1         \\
\hspace{0.2em} Undisclosed                                 & 4   & 3         \\
\textbf{Education}                          &     &           \\
\midrule
\hspace{0.2em} High school                                 & 73  & 14        \\
\hspace{0.2em} College                                     & 114 & 37        \\
\hspace{0.2em} Undergraduate                               & 254 & 78        \\
\hspace{0.2em} Postgraduate                                & 118 & 30        \\
\hspace{0.2em} Doctoral level                              & 12  & 3         \\
\hspace{0.2em} Undisclosed                                 & 4   & 3         \\
\textbf{Income (based on national average)} &     &           \\
\midrule
\hspace{0.2em} Below average                               & 161 & 33        \\
\hspace{0.2em} Average                                     & 237 & 88        \\
\hspace{0.2em} Above average                               & 151 & 38        \\
\hspace{0.2em} Undisclosed                                 & 26  & 6        
\end{tabular}
\end{table}

\subsubsection{Data analysis}
On analysing this data, we identified 165 (28.7\% of total) participants who, during the online study, typed at least one message that our proactive moderation system detected and flagged as containing toxicity. In total, 203 individual messages were flagged as toxic and triggered a moderation event at least once, and a total of 225 events occurred when including our control group. These events were approximately evenly distributed across our study conditions, with a mean of 25 events occurring across each condition (SD=13.06). Of the participants who had a message flagged for moderation, 83.03\% (137) had one message flagged, 12.12\% (20) had two messages flagged, 4.24\% (7) had three messages flagged, and one participant (0.61\%) had five messages flagged. The quantitative analysis reported here uses the 203 moderation prompt events dataset. Where a single message triggered a moderation prompt more than once (e.g., a participant typed a toxic message, then revised it to be non-toxic, and revised again to be toxic), we use the first moderation prompt event only within our analysis. 

We included two dependent variables within our analysis. The continuous scale of \textit{Toxicity level} was calculated on the typed messages using Perspective API. We captured the toxicity level twice, firstly when the message was flagged (pre), and secondly when it was sent (post). As \textit{toxicity level} is repeated, we performed a repeated measures ANOVA. Before using this test, we checked assumptions. We screened for outliers with Mahalanobis distance, finding no outliers within the data. We found slight violations of normality within the data. However, ANOVA is robust against violation of normality where larger sample sizes exist. As our between-subject variable of toxicity consists of two levels, sphericity necessarily holds. Finally, to check for equality of covariance matrices we used Box’s test, which was non-significant (p > .05), meaning the covariance matrices of the dependent variables are equal across our groups. Finally, an a priori power analysis to estimate sample size resulted in a sample recommendation of 138 (within-between interaction) where $\alpha=0.05, 1-\beta=0.8, f=0.1$). As our assumptions and sample size were met for repeated measures ANOVA, we proceeded with our analysis. 

The second dependent variable included within our analysis was~\textit{revision status} (i.e., whether a message was, or was not revised). We determined~\textit{revision status} by comparing the pre and post-moderation scores. If the change between scores was 0, we define the event as not revised. If > 0, we define the event as revised. As this is a binary variable, we performed a binary logistic regression, with \textit{timing}, \textit{presentation}, and \textit{friction} as our predictor variables. As our dependent variable is dichotomous, our independent variables are categorical, and our observations are independent, we satisfy our test assumptions and proceed with our analysis.

%To test our hypothesis, we performed an analysis of covariance (ANCOVA) with the pre-toxicity score (initial toxicity score prior to intervention) as a covariate, and our post-toxicity score (the score of the submitted message) as our dependent variable. We included three between-subject factors: timing (send vs typing), friction (low vs high), and presentation (abstract vs explicit) which created 8 groups. Before running our analysis, we tested the main assumptions for ANCOVA. We screened for outliers with Mahalanobis distance, finding no outliers within the data. We found slight violations of normality within the data. However, ANCOVA is robust against violation of normality where larger sample sizes exist. To check for homogeneity of variances we used Levene’s test, finding it to be not statistically significant (F(7,195) = 1.33, p = > 0.5). Next, we checked for the homogeneity of regression slopes. To do this, we ran an ANCOVA to check for the statistical significance of our covariate by experimental conditions (timing, friction, presentation). Our covariate by timing ($F(1, 215) = .11$, $p > .05$), friction ($F(1, 215) = 1.90$, $p > .05$), and presentation ($F(1, 215) = 0.92$, $p > .05$) treatment interactions were not statistically significant, meaning the homogeneity of regression slopes assumption is met. Satisfied that our assumptions for ANCOVA were met, we proceeded with the analysis.

\subsection{Study 2 (Interview study)}
To gain a more in-depth understanding of how users perceived and experienced the different proactive moderation prompt designs, we ran semi-structured interviews with participants across our experimental conditions. 

\subsubsection{Participant recruitment and interview protocol}
From our sub-sample of those that triggered a moderation event, we performed purposeful sampling to  capture insights from a diverse range of participants. This included participants across the eight design groups (see: Table~\ref{tab:demographics_interview}), participants who revised their message to be more toxic, less toxic, and those who chose not to revise. In total, we interviewed 17 participants (see: Table~\ref{tab:demographics_interview}) between 19 and 22 of December, with 1 being conducted on the 3rd of January 2023. We ran the interviews soon after the online study to enhance memory recall of the online study experiment. The first author conducted all interviews anonymously via Zoom web. Only audio was recorded, alongside an automatic transcription of the conversation. Interviews continued until (1) a sample from all intervention groups was obtained, (2) no further insights were being obtained (3) the first author evaluated that the data collected would allow for the research questions to be adequately addressed. Interviews lasted an average of 47 minutes (34m-63m). All interviews were semi-structured which provided interview structure whilst giving the flexibility to explore interesting avenues of discussion. The first author kept handwritten notes during each interview consisting of insights observed during the interviews. These acted to support the interview and the subsequent data analysis. 

% Please add the following required packages to your document preamble:
% \usepackage{graphicx}
\begin{table}[]
\centering
    \caption{Demographics of our interview participants ($N=17$), and the intervention they used during the initial online study. We also provide a summary of our interview participants' prior experiences of being moderated by a proactive moderation system.}
    \label{tab:demographics_interview}
\resizebox{\textwidth}{!}{%
\begin{tabular}{lllllllll}
\toprule
\multicolumn{1}{c}{-} & \multicolumn{5}{c}{\textbf{Demographic Information}}                                                                                                                                   & \multicolumn{3}{c}{\textbf{Intervention Information}}   \\
\textbf{PID}          & \textbf{Gender} & \textbf{Age} & \textbf{Education}   & \textbf{Income}        & \textbf{\begin{tabular}[c]{@{}l@{}}Prior Experience\end{tabular}} & \textbf{Timing} & \textbf{Friction} & \textbf{Presentation} \\
\midrule
P01                   & Male            & 35 - 44      & High school          & Below national average & No                                                                                                    & Typing          & High              & Abstract            \\
P02                   & Male            & 35 - 44      & College              & National average       & No                                                                                                    & Send            & Low               & Explicit             \\
P03                   & Male            & 45 - 54      & College              & National average       & Yes (Twitter)                                                                                         & Typing          & Low               & Explicit             \\
P04                   & Male            & 24 - 34      & Undergraduate Degree & Above national average & No                                                                                                    & Send            & High              & Abstract            \\
P05                   & Male            & 24 - 34      & Undergraduate Degree & Above national average & No                                                                                                    & Typing          & Low               & Explicit             \\
P06                   & Male            & 35 - 44      & Undergraduate Degree & Above national average & Yes (Twitter)                                                                                         & Typing          & High              & Explicit             \\
P07                   & Female          & 65+          & Postgraduate Degree  & Undisclosed            & No                                                                                                    & Send            & Low               & Abstract            \\
P08                   & Male            & 35 - 44      & Undergraduate Degree & Above national average & Yes (Twitter)                                                                                         & Typing          & Low               & Abstract            \\
P09                   & Female          & 35 - 44      & Undergraduate Degree & Above national average & No                                                                                                    & Send            & Low               & Abstract            \\
P10                   & Male            & 45 - 54      & High school          & Below national average & Yes (Web Forum)                                                                                       & Typing          & High              & Explicit             \\
P11                   & Male            & 35 - 44      & Undergraduate Degree & Above national average & No                                                                                                    & Typing          & Low               & Abstract            \\
P12                   & Male            & 24 - 34      & Undergraduate Degree & Above national average & No                                                                                                    & Send            & Low               & Abstract            \\
P13                   & Male            & 35 - 44      & Postgraduate Degree  & Undisclosed            & Yes (Twitch)                                                                                          & Typing          & High              & Explicit             \\
P14                   & Female          & Under 24     & College              & Average                & Yes (Dating App)                                                                                      & Typing          & Low               & Abstract            \\
P15                   & Female          & 24 - 34      & College              & Average                & No                                                                                                    & Typing          & Low               & Abstract            \\
P16                   & Female          & 35 - 44      & Undergraduate Degree & Above national average & No                                                                                                    & Send            & Low               & Explicit             \\
P17                   & Female          & 35 - 44      & Undergraduate Degree & Above national average & No                                                                                                    & Send            & High              & Explicit            
\\\bottomrule
\end{tabular}%
}
\Description{Demographics and intervention assignment}
\end{table}

The full interview script can be viewed in the supplementary material. Interviews started by asking participants about their prior experience with proactive moderation systems (see Table~\ref{tab:demographics_interview}). We then explored participants' general views related to this form of moderation both as an independent keyboard, and a tool built into a social networking platform. We probed people's concerns, how they would feel if they were using the system, and how it might impact how they perceived other people's communications. We then asked questions related to the online study to understand what participants remembered from the moderation prompts they received, why they thought they had received them, and what they believed the intention of the prompt was. After this, we presented them with the data they generated during the online study including draft messages that triggered prompts, a visual of the prompt, revised messages, and sent messages. We explored their messages to understand what they thought had triggered the prompt, whether it was appropriate to have been triggered, how they felt and reacted when receiving the prompt, and why they had responded in the way they did (e.g., revise, not revise). Following this, we explored the other design conditions with participants to understand their views of the other design factors. We then asked a series of questions related to classification errors (e.g., false positives) and how they would feel and respond if they experienced a classification error. Finally, during the online study, we asked each participant to describe a personal experience in which they had sent an inappropriate message. We read back the text (if) they submitted and asked them how a moderation system like the one used, may have affected their experience. 

\subsubsection{Interview data analysis}
All automatically transcribed interviews were checked and corrected by the first author. Transcripts were cleaned to remove unnecessary words to improve readability (e.g., repeated words). The first and second authors worked collaboratively to code and analyse the data, applying an iterative inductive coding method based on the stages outlined by Braun and Clarke~\cite{braun2006using}. Practically, this involved an initial round of open coding where the first author applied a complete coding approach where all data was coded in Nvivo. After half of the transcripts were coded (randomly selected), the first and second authors met to discuss the codes, reading examples from the transcripts. These codes were then moved into an online collaborative whiteboard (Miro) to support the development of themes from the codes. The first and second authors met again to discuss the sorting of codes into themes, further revising the themes. Once agreed, the codes were moved back into Nvivo, where each transcript was coded deductively using the developed codebook.

\section{Quantitative Results}
We first report on our quantitative analysis of proactive moderation events to provide a broader understanding of how participants responded after being prompted, and how the three design variables (timing, friction, presentation) impacted on levels of toxicity of sent messages. We ran a repeated measures ANOVA  where the between-subject factors (timing, friction, presentation) and within-subject (repeated measure) factor of toxicity (pre/post) were included within the model. Message toxicity level is the dependent variable, calculated using perspective API~\cite{perspectiveWeb}. 

\subsection{Descriptive statistics}
\begin{figure}
\centering
\begin{minipage}[t]{\dimexpr.5\textwidth-1em}
  \centering
  \includegraphics[width=1\linewidth]{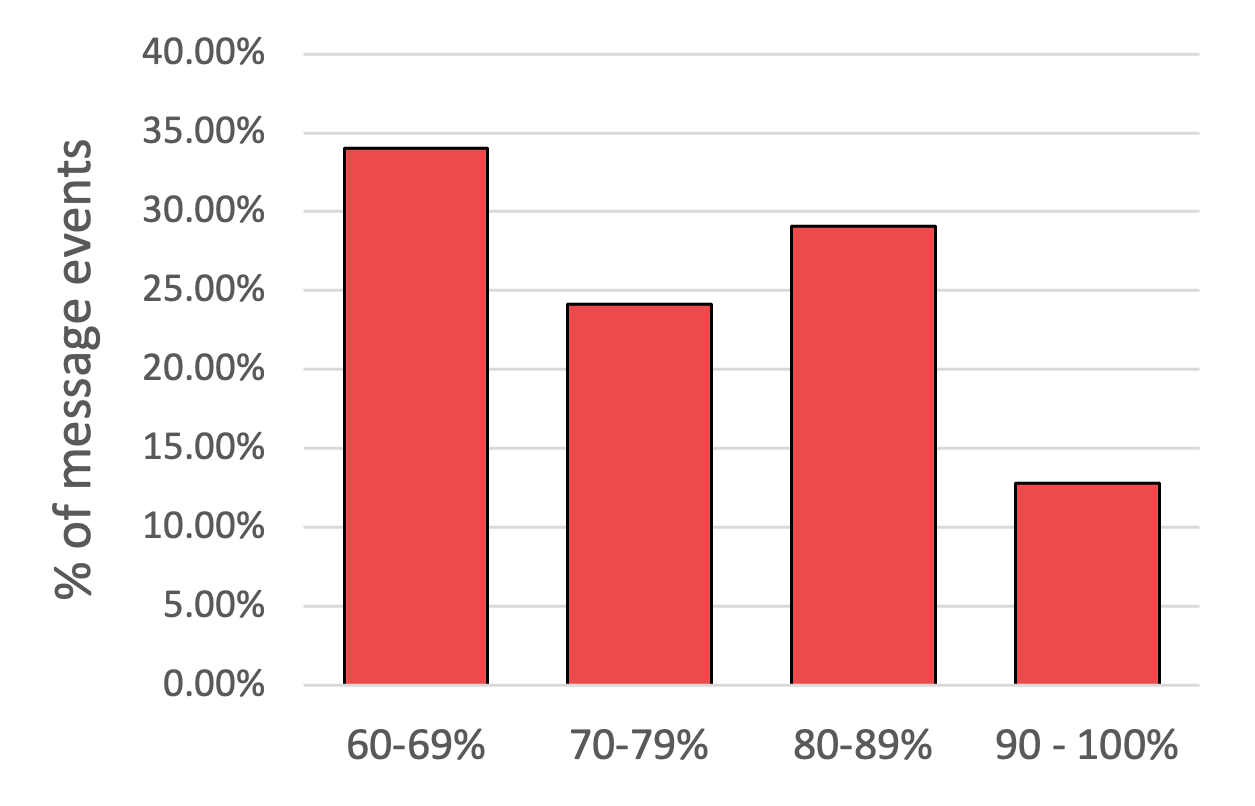}
  \captionof{figure}{Graph showing the \% of message events by toxicity level. This shows that most messages were of a toxicity level between 60\% and \~89\%, with fewer at more than 90\%. }
  \label{fig:event_by_level}
\end{minipage}\hfill
\begin{minipage}[t]{\dimexpr.5\textwidth-1em}
  \centering
  \includegraphics[width=1\linewidth]{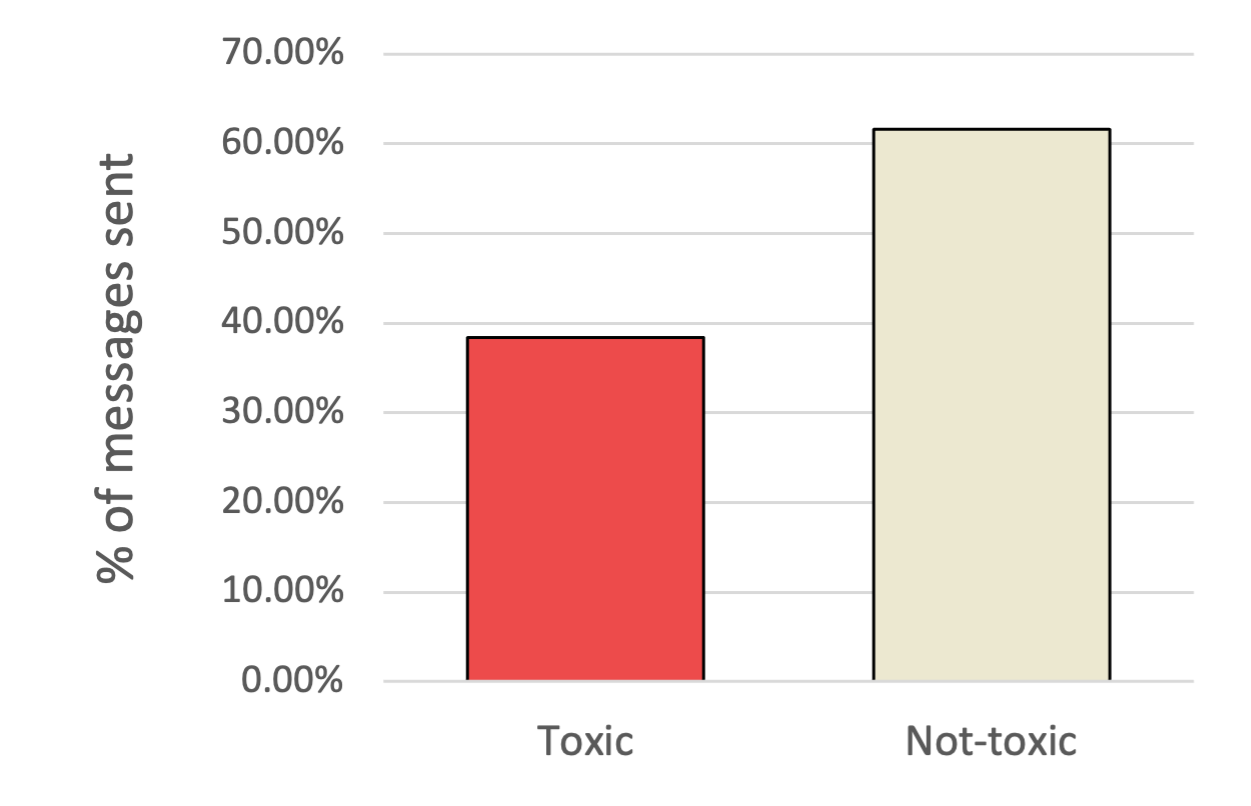}
  \captionof{figure}{Graph showing the \% of messages that were toxic and non-toxic at the point of sending. This shows that \~60\% of messages went from toxic to not-toxic as a result of the moderation prompt.}
  \label{fig:sentstate}
\end{minipage}
\end{figure}

Of the messages that triggered a moderation event,  most received a toxicity score between 60\% - 89\%, with fewer being triggered at the higher end of > 90\% (see: Fig~\ref{fig:event_by_level}). Of the messages that at some point received a toxicity score of more than 60\%, most were revised to be less than 60\% (61.58\%, 125) (see: Fig~\ref{fig:sentstate}). Our sample of moderation events includes more men (60.61\%, 100) than women (36.97\%, 61). Moreover, the mean toxicity score was higher for messages triggered by men ($M = .77$) compared to women ($M = .75$). To explore whether there was a statistically significant difference, we ran an independent sample t-test to compare pre and post-moderation toxicity of messages between men and women. There was no significant difference in pre-moderation toxicity between men ($M = .77$, $SE = .02$) and women ($M = .75$, $SE = .02$); $t(159) = -1.58$, $p > .05$. There was also no significant difference in post-moderation toxicity between men ($M = .44$, $SE = .05$) and women ($M = .45$, $SE = .05$); $t(159) = 0.16$, $p > .05$. Our sample contained more moderation events that occurred within a WhatsApp scenario (81.77\%, 139), than in a Twitter scenario (18.23\%, 26). To compare the pre and post-moderation toxicity levels of messages across the two platforms, we ran an independent sample t-test finding no statistically significant difference in pre-moderation toxicity levels ($M = .04$, $SE = .02$); $t(163) = 1.86$, $p > .05$, or post-moderation levels ($M = .06$, $SE = .06$); $t(163) = .31$, $p > .05$. Given the small sample of Twitter based events collected, further quantitative analysis combines these platforms. 

\subsection{Message toxicity level before and after moderation}
\begin{figure}[t]
\centering
\begin{subfigure}{.33\textwidth}
  \centering
  \includegraphics[width=1\linewidth]{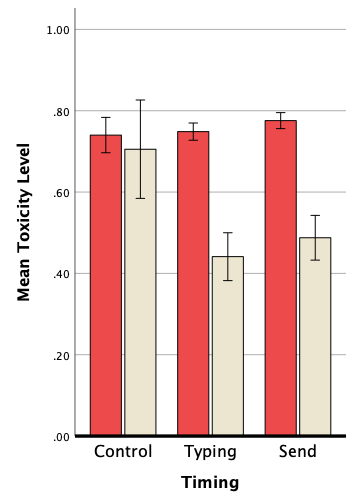}
  \caption{Intervention timing}
  \label{fig:toxicityVsVariables1}
\end{subfigure}%
\begin{subfigure}{.33\textwidth}
  \centering
  \includegraphics[width=1\linewidth]{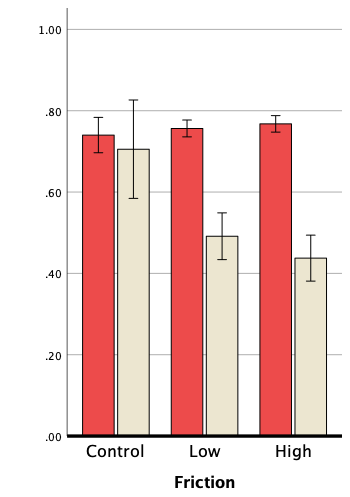}
  \caption{Intervention friction}
  \label{fig:toxicityVsVariables2}
\end{subfigure}
\begin{subfigure}{.33\textwidth}
  \centering
  \includegraphics[width=1\linewidth]{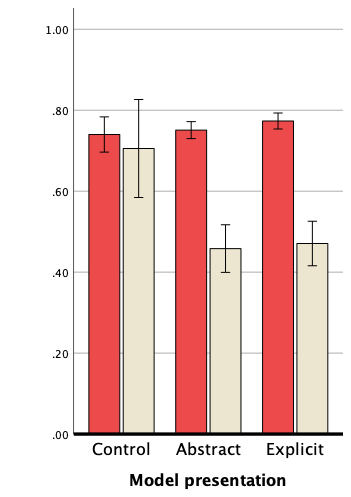}
  \caption{Model presentation}
  \label{fig:toxicityVsVariables3}
\end{subfigure}
\caption{Toxicity levels pre-moderation (red) vs post-moderation (cream) across the three experimental conditions. Error bars at 95\% CI.}
\label{fig:toxicityVsVariables}
\end{figure}

We explored the impact of the moderation intervention against the between-subject variable of toxicity (pre/post). We found a statistically significant main effect on toxicity ($F(1, 216) = 105.51$, $p < .001$) with pre-moderated messages being more toxic than post-moderated messages, $M = .27$, $SE = .02$, $p < .001$, 95\% CI [.23, .31]. To investigate the impact of the intervention against our control, we explored differences in toxicity (pre/post) in our control group finding no statistically significant difference, $M = .04$, $SE = .06$, $p > .05$, while finding a statistically significant difference for all other conditions ($p < .001$) (see: Fig~\ref{fig:toxicityVsVariables}). Next, we investigated the interaction effects between toxicity and our between-subject variables. We found no statistically significant effect between toxicity and timing ($F(1, 216) = .22$, $p > .05$), toxicity and presentation ($F(1, 216) = .05$, $p > .05$) and toxicity and friction ($F(1, 216) = 2.39$, $p > .05$). Additionally, we found no statistically significant interaction effect between toxicity, timing, and presentation ($F(1, 216) = 0.04$, $p > .05$); and toxicity, presentation, and friction ($F(1, 216) = .24$, $p > .05$). However, an effect was found between toxicity, timing and friction ($F(1, 216) = 5.20$, $p < .05$). 

To investigate this finding further, we ran pairwise comparisons. For participants exposed to moderation on~\textit{typing}, the level of \textit{friction} (low vs high) had no effect on post-moderation message toxicity ($p > .5$, $M = -.05$, $SE = .06$). However, for those prompted on~\textit{send}, \textit{friction} (low vs high) did have a statistically significant effect ($p < 0.5$, $M = .15$, $SE = .06$, 95\% CI [.04, .26]) on post-moderation message toxicity. For participants in a~\textit{high friction} group, the~\textit{timing} (typing vs send) of the intervention had no statistically significant effect on post-moderation message toxicity ($p > .05$, $M = -.03$, $SE = .02$). However, for those in a \textit{low friction} group, intervention~\textit{timing} (typing vs send) did have a statistically significant effect ($p < 0.5$, $M = -.15$, $SE = .06$, 95\% CI [-.26, -.03]). Finally, we investigated the interaction between~\textit{toxicity}, ~\textit{timing},~\textit{presentation}, and~\textit{friction}, finding no statistically significant effect ($F(1, 216) = 2.30$, $p > .05$).

\subsection{Message revision}

Next, we investigate how our independent variables impact the likelihood of a moderated message being revised, by analysing our binary dependent variable~\textit{revision status}. The results from this indicate that~\textit{timing} and~\textit{friction} are significant predictors of~\textit{revision status} [Chi-Square=61.94, df=3 and $p < 0.001$]. The other predictor ~\textit{presentation} was not significant. All three of the predictor variables explained 40.60\% of the variability of~\textit{revision status}. Variables~\textit{timing} and~\textit{friction} are significant at the 5\% level [~\textit{timing} Wald = 21.64, $p < .001$;~\textit{friction} Wald = 11.42, $p < .001$]. The odds ratio for~\textit{timing} (typing vs send) is .03 (95\% CI [.00, .13]) and for~\textit{friction} (low vs high) is 4.09 (95\% CI [1.81, 9.25]. This shows that messages were more likely to be revised where moderation was implemented~\textit{on send}, and where~\textit{friction} was higher. The model correctly predicts 83.6\% of cases where a message revision is made and 68.2\% of cases where no revision is made. The overall correct prediction rate is 80.3\%.

\section{Qualitative Findings}
While the quantitative findings provide useful insights, we wanted to understand the qualitative experiences and perceptions of proactive moderation systems (RQ3) to surface insights that may otherwise become lost within the quantitative statistics. 
We report four themes developed from our interview analysis. We asked participants about their experiences using the study app keyboard, as well as their prior experiences with proactive moderation systems. While most responses relate to the use of the study app keyboard, where responses relate to prior experiences, we have explicitly stated this. The first theme explores the perceived purpose of the system, and its boundaries (i.e. to what extent should the system moderate content). The second explores the perceived and experienced limitations of the system and how these impact user experience. The third highlights user rights that the system would need to be developed around to be acceptable. Finally, we explore the impact the system can have on conversations and relationships, and on impression management. Throughout the findings, we have~\textit{italicised} mentions of intervention factors (e.g.,~\textit{on send} as a timing factor), and reported abbreviated study conditions against each participant: Timing (~\textsuperscript{Type} vs ~\textsuperscript{Send}), Friction (~\textsuperscript{High} vs ~\textsuperscript{Low}), Presentation (Abstract (~\textsuperscript{Ab}), vs Explicit (~\textsuperscript{Ex})). 

\subsection{Perceived purpose of the system and its boundaries}\label{purpose}
\subsubsection{Improving the safety of online spaces}
Participants highlighted how the system could help to increase the safety of different online spaces by acting to prevent harmful content from being sent. They recognised a need for the system having experienced or witnessed problematic conversations online. \nineA said she would~\say{feel a little bit more protected from it [harmful content]} if the system were to be deployed, and \sixteenA highlighted how she would~\say{get the benefits because I wouldn't see things that offend me as much}. 
%In line with our quantitative findings, we find gender differences in our qualitative results. From our sample, women mentioned feeling safer if the system was to be implemented, more than men. However, the male participants did still see the safety benefits but often did not feel as though they themselves would benefit. For example, \threeA said \say{I don't feel as if I need to be protected, but if it's going to protect others who might be more vulnerable, then I’m all for it}. 

\subsubsection{Creating moments for reflection}
All participants recognised the benefit the system could bring in creating moments for reflecting to help reduce toxic content. Even where messages were not revised, participants often reflected on their messages, and how they made more~\say{conscious decision[s]} on how to proceed. Supporting our quantitative findings,~\textit{friction} helped to increase reflection, with \fourA saying ~\say{I would have instantly pressed send. With the friction, it does give you a chance, that 30s or so is good enough to have a little think about what you said}. \fiveA described how they felt guilt on receiving the moderation prompt, saying \say{you feel a bit of guilt about what you're about to write, so you instinctively review it and check it and see how it sounds}. \oneA described this process, saying:~\say{it's kind of taking the emotion out of your sort of responses sometimes, or making you at least think about it}. %While removing or reducing the emotion from the conversation may help to reduce the harm an interaction can cause, in section~\ref{impressionmanagement} we highlight how this may have negative implications on communication and impression management.

\subsubsection{Education and awareness through moderation}
Participants described how the process of reflecting on messages could help educate and increase people's social awareness. Participants spoke about using terms or language without realising how inappropriate it was, and why it was deemed inappropriate. For example, \sixteenA said~\say{if you've used something inappropriately and you’re not aware of it, maybe it’s telling you, it’s giving you more information on what the appropriate language is or whatever}. \elevenA also identified a potential benefit of educating through moderation, saying~\say{if it helps people be educated to find different ways of saying things, so to communicate better and not be so offensive, then that's also good}.

\subsubsection{Content moderation and restriction}
When participants were asked about the type of content that the system should moderate, responses ranged from content containing coarse language to content that could lead to physical harm. Participants who voiced concerns over how the system could encroach on their freedom of speech were in favour of moderation being restricted to more extreme content such as content that could result in someone being~\say{injured, hurt, [..] murdered, whatever it may be}~\six. There were particular concerns raised around children and the need to protect them from harm. When \fourA was asked about the type of content that he felt should be flagged by the system, he said: \say{sexual, racist, obviously thinking like people grooming kids online and stuff}. ~\threeA made their evaluation by drawing parallels with offline laws, restricting moderation to content that ~\say{potentially would be deemed against the law}.  

Racial and sexual language was commonly mentioned by participants as requiring moderation, and this included harassment behaviours. \fourteenA stated that she would consider~\say{anything of like a sexual harassment kind of nature} to be moderated, and similarly, \tenA felt that~\say{racist remarks, sexist remarks, online bullying, [..] prejudice comments} should be moderated by this system. There was also a view that the targeted nature of the message should be taken into account. \fiveA said~\say{the fact that it's `you' [means] it's directed at someone} which resulted in the message being perceived as more harmful. Some felt that foul language should be subjected to moderation by the system. \eightA said the system should detect~\say{course language, you know, abusive name calling}. Other participants wanted bullying-related language to be subject to moderation.

We also explored perceptions of embedding restrictions into the system. Some participants used a proactive moderation system that incorporated a~\textit{friction}, and so had already experienced a type of restriction. However, we were interested to understand how far a system should be permitted to go. For instance, whether users should be blocked from sending inappropriate content, and if so when this would be appropriate. When we explored the idea of adding a friction with \threeA, he said~\say{yes, guide me, advise me but then to restrict me, I think might put me off}. Yet, when he spoke of blocking inappropriate messages, he was more favourable where the content~\say{was deemed illegal}; yet concerns were raised around how accurate the system would be at determining the legality of content. 

\subsection{Limitations of proactive moderation systems}
\subsubsection{Accuracy of the system and its impact on fairness}
A perceived lack of contextual understanding that the system would have was a concern amongst many, not just within sentences being typed but also around the broader communication and relationship context. In particular, this concern was raised around conversations with close ties such as friends and family where there are social norms related to in-group dynamics (e.g., language, knowledge, and behaviours). \twoA highlighted a recent conversation he had with a friend who had been cheated on, saying:~\say{there was a fair bit of profanity [that] wasn't actually aimed specifically at each other, it was more like me showing empathy and sympathy to the situation}. Similarly, \oneA was concerned that when using \say{certain words but not in like an attacking way, just to sort of make a point}, the system would flag it as~\say{harmful or bullying, or something when it isn't}. A perceived limitation of the system was its inability to detect humour; a concern raised when participants again spoke about communicating with close ties. For example, \elevenA said~\say{I know people in friendships can say things that are, yeah, meant in humour, and maybe the algorithm wouldn't know the difference}. Despite these concerns, participants recognise that~\say{nothing is 100\% accurate} \three and that the system would make mistakes. For participants who were positive towards proactive moderation, this was not a barrier to use. \seventeenA highlights this, saying~\say{I'd rather have it and have it there for when it needs to be used and have it pop up sometimes when it doesn't need to, [..] than not have it.}

The perceived fairness of the systems was often entwined with issues of accuracy and the systems inability to detect more subtle forms of toxicity. Some participants highlighted how their messages were flagged because they contained coarse language, while other messages in the thread were of an abusive nature but were not so explicitly abusive, and so may not have been flagged. When asked about his message and whether it should have been flagged, \sixA said: "I can understand why it did it. I don't think it was appropriate, no". When asked about the fairness of the prompt, he said ``I’ve been on the receiving end of [..] all out abuse, but it was because they didn't use naughty words they just got to carry on saying whatever they liked''. 

\subsubsection{Communication efficiency}
Participants spoke of the additional effort required to revise messages, and how this would slow down their interactions and conversations. Although the system aims to slow down interactions when toxic content is detected, there were concerns over how disruptive this may be in different contexts.~\nineA said~\say{if you are, say, having an argument with a friend or family member it would be quite irritating if you were trying to kind of argue your point and kept getting these pop-ups all the time}. Some participants expressed a concern that highly disruptive prompts could have an undesirable effect,~\say{if it kept coming up while you were typing, you might get even more annoyed}~\seven.  

The \textit{timing} of when prompts occurred appeared to impact revision speed. Participants using~\textit{on typing} interventions spoke of their ability to make changes~\say{there and then}~\two with \threeA saying \say{it was good that at the point of it determining that was inappropriate language or context, that it did warn me at that time [and] didn't wait till the end}. The \textit{on typing} timing appeared to make it easier to understand which part of the message was considered inappropriate by the model. In contrast,~\textit{on send} interventions required an evaluation of the entire message, slowing down the interaction; a finding consistent with our quantitative results. \twoA said~\say{I just think it'd be a pain. I'd be more likely to revise something as I’m writing}. 
%\fiveA said:~\say{to have it with you in real-time was definitely better than having to sort of send it away and wait for an answer. It was much more [..] interactive, it was, I don't want to wait for things basically}. 
Discussions related to communication efficiency were primarily focused on communication speed. However, some participants spoke of the need for both speed and accuracy, \twelveA said~\say{I think in communication, accuracy matters more than speed}. %We explore the issue of accuracy further when exploring the role these systems have in relation to impression management behaviours (section:~\ref{impressionmanagement}). 
Speed appeared more of a concern within instant messaging applications (e.g., WhatsApp), when compared to social media platforms (e.g., Twitter) due to the nature of the interactions. One participant highlighted this explicitly, saying:~\say{Twitter, it's [..] slow. WhatsApp is more like a, you exchange a message and every time you have this [..] warning to say ‘Yes, okay’, ‘Yes, I did this, yes, okay’, and every time it's slowing the conversation, it's not nice} \thirteen.

\subsubsection{Overriding moderation prompts}
Where the model was inaccurate due to a lack of contextual understanding, participants described how they read the moderation prompt but chose not to revise their message before to sending it. %Reasons for this were varied and some related to the design of the moderation system. 
Supporting our quantitative findings,~\textit{timing} appeared to impact on whether a participant would revise, and this was linked to the effort required to revise a message after it has been completed, and the level of commitment the participant had over their drafted message. \nineA said~\say{it's like you've already committed to it}. While participants often felt as though the moderation was justified, where they reported being satisfied with what they had written, they proceeded to send regardless. \nineA spoke about one of the messages he chose not to revise, saying:~\say{I was just thinking they're a cheeky so and so, and I would not stand for that if that was my family talking about my partner like that [..], I was thinking about [revising] but I was quite happy with my response}. 

\subsection{Respecting the rights of users}

\subsubsection{Intrusiveness of the moderation approach} 
There were concerns over the system being overly intrusive and at risk of becoming a surveillance tool. \sixA felt the system was \say{a little bit big brothery} in the way it worked. Others agreed that it felt \say{controlling and almost sort of nannying} \eight and could curtail freedom of speech and result in feeling restricted. \thirteenA said~\say{I didn't feel that I was free in that kind of platform to express what I really feel so I don’t I think it's nice, it's not a nice feeling}. However, these concerns were not universally shared. While \fifteenA recognised the concern saying~\say{other people might worry that their free speech is being [..] interfered with}, when asked whether they would personally, they said~\say{not personally, no} reporting a general lack of concern related to issues of online privacy and security.

One way in which the intrusive nature of the system was felt was through people's anticipation of the moderation system. While this is also an aim of the system (i.e., to improve people's communication behaviour) it also raised concerns related to the curtailing of freedoms. \tenA said~\say{I'm pretty sure I tried to use it to not have it pop up again but I wouldn't say that's necessarily a good thing}. When asked why, he said the system was acting~\say{like judge, jury, and executioner} and did not see~\say{why I should water myself down just to be able to get my point across}. This concern was more apparent for those being moderated \textit{on typing} as this further restricted their ability to write out their message in full before being moderated. \seventeenA said~\say{doing it at the time of sending it lets you still get [..] out all your frustrations and then you can just change it after}. While~\textit{on typing} prompts may require less effort to revise they can restrict users from writing out a response to express their anger or frustration. We refer to these types of messages as `phantom messages' as they are written, but never sent. 

\subsubsection{Privacy and transparency}
Throughout, participants raised concerns over what data would be collected, how long it would be stored, where it would be stored, how it would be processed and by whom. For some, there was an acceptance that privacy would be compromised to benefit from the system, comparing it to business models of other platforms and technologies. \oneA said:~\say{it's hard to kind of understand where all that data is going, really. So [..] you understand it's kind of like a cost of doing business when you go online}. There was more of a concern around data collection of messages within instant messaging applications where data is more private, compared to more public social platforms like Twitter. %The type of information being shared across instant messaging platforms was seen as more sensitive. 
For example, when \twoA was speaking about his data privacy concerns, he said~\say{I would quite happily send my [bank] card details to my mother, I would not be happy if that was being recorded}. Again, the design of the system appeared to impact privacy concerns with those being moderated~\textit{on typing} concerned that their `phantom messages' would be recorded. \fourA said~\say{I guess it's quite worrying, [..] especially if it's something that you're not going to actually send}. 

There were also concerns about the transparency and fairness of the algorithm supporting the moderation system. The main transparency concerns related to who was making the decision, and who was deciding what is and is not appropriate. When \sixA was asked about a moderation prompt he received during the study that presented him an~\textit{explicit} \% score he said: \say{According to who? According to what? According to Guardian readers? Well bloody hell, according to The Times readers? According to Sun readers?} Here, the mentions of various newspapers suggest a concern of political bias within the system. 

\subsection{Impact on communication and relationships}\label{impressionmanagement}

\subsubsection{Impression management}
%In section~\ref{purpose} we report how our participants see this moderation approach as supporting reflection to improve the quality of discourse and reduce instances of harm. However, 
As a sender, participants identified how the system could help them better manage the impression they ``give off" online, especially where they are ``performing" to a larger audience, such as Twitter. \twelveA spoke about this, saying:\say{whenever you're talking to someone you're always showing them a view of yourself [..] the sender might have a chance to think about how they come off}. Similarly, \fifteenA spoke of the system helping to reduce the risk of sending a message in the~\say{spur of the moment} suggesting that it \say{might just make you take a step back and just rethink}. \fifteenA and other participants describe how reflective prompts help to reduce the levels of emotion within the conversation, to avoid people sending messages that they may later regret. Yet, \oneA questioned the impact this could have on conversations and relationships, especially where it limits a person's ability to portray their ``true" self. He said: \say{maybe you could have not responded as emotionally, and you know it's better. But I think in general that's who we are, that's who I am, that's being a human}.

\subsubsection{Impacting on how people perceive others}
Where the system was seen as mediating communications, there were concerns over how people would perceive received messages. \seventeenA was particularly concerned about this, saying: \say{you're having a disagreement and they send you a message, and all of a sudden it's been very kind of polite and they've stopped being argumentative, you might be thinking~\say{well, I don't think that you're actually feeling that way or [..] that's what you actually want[ed] to send to me, it's just [..] what you're [..] sending as a result of the keyboard diffusing the situation, even if you don't want it diffused}}. In contrast to this, participants also spoke of how the system may increase the perceived intentionality behind inappropriate messages received, as they may become implicitly aware that the sender had been prompted to revise, and ignored the prompt. \fiveA said: \say{I would probably view the person or the message a bit more, a bit more deliberately because they've had the chance to review the automated message, and they nonetheless sent it so it would, it would kind of possibly be worse than the impact of the message}. However, other participants spoke of how they may simply~\say{assume that they're ignoring it [the prompt]}. %\oneA used a keyboard that moderated~\textit{on typing} with~\textit{low friction} which may have made ignoring moderation prompts more likely as a result of the less intrusive nature of this design.

\section{Discussion}
%~\textit{Presentation} of the model's output had no impact on the effectiveness of the moderation prompt in reducing message toxicity. %However, our qualitative findings suggest that explicit presentation may result in users gamifiying the system either for playful reasons or with more malicious intent. 
%Our findings also suggest more explicit presentation can help users to test the system and its boundaries with the intention of circumventing detection. While this type of behaviour may be exhibited by only a small percentage of users, it may still result in harm. 

This research adds weight to the growing body of work showing how proactive moderation can be effective at reducing message toxicity~\cite{seering2019moderator, chang2022thread,van2017thinking,jones2012reflective,vincent2020twitter, katsaros2022reconsidering,wang2013privacy}. Our findings support prior work~\cite{wang2013privacy,wang2014field,masrani2023slowing} that shows how people can become frustrated when friction is built into communication interactions; and while the friction within our platform was only present on the detection of toxic content, it was still a cause of annoyance. We found this to be especially problematic within private IM platforms due to the speed people exchange messages compared to public social media platforms. Moreover, as prior work shows people often make bad communication decisions when they are in a negative emotional state~\cite{wang2011regretted,sleeper2013,cheng2017anyone}, we find concerns around how disruptive moderation prompts could irritate users who are already in a volatile state. Even though messages were more likely to be revised where moderation was implemented~\textit{on send} and when ~\textit{friction} was high, we found~\textit{low friction} and~\textit{on typing} interventions to be more effective at reducing message toxicity. \textit{On typing} prompts also reduced the perceived time it took participants to revise their message when toxicity was detected and allowed them to better understand the point at which their message became toxic. However, toxicity models typically make predictions of how likely a piece of text is to be toxic, as opposed to how likely a message is to become toxic once complete. Where detection occurs on typing, a partial message may be considered toxic, while the complete (and originally intended) message could be considered non-toxic, thus incorrectly flagging messages. ~\textit{Presentation} of the model's output had no impact on the effectiveness of the moderation prompt in reducing message toxicity.

\subsection{Private vs Public Platform Moderation} %7.1
Prior research into proactive content moderation focuses on public platforms, rather than private platforms like chat apps (e.g., WhatsApp)~\cite{katsaros2022reconsidering,chang2022thread,cheng2017anyone}. In taking a cross-platform approach, we identified distinctions between private and public platform implementation due to differences in the way people perceive and interact across these spaces. As communication with close ties is more common within private channels, these spaces rely more heavily on in-group social norms around language, knowledge and behaviour. These more intimate conversations were viewed differently from public conversations, resulting in heightened concerns about how conversations within these intimate spaces were being processed. When developing proactive moderation systems for private spaces, social norms that may result in non-toxic language being flagged as toxic need to be considered. The often contextual and cultural nature of language means the language used by one group may be interpreted very differently from that used by another; in some cases, the flagging of content as toxic may in and of itself be offensive. Moreover, from an interaction perspective, we highlight the different communication styles inherent across the two platform types, with private chat apps encouraging faster communications. Where~\textit{frictions} is applied to the design, this becomes more distributive within private platforms. The other distinction we identified relates to the privacy of communications and perceptions of surveillance. Proactive moderation on public platforms does not require scanning of content on the user's device (often referred to as client-side scanning (or `CSS')), as the content is freely available on the platform server. CSS is required where proactive content moderation is performed on private platforms where messages are end-to-end encrypted as the content is only visible on the client device (either sender or receiver). Researchers have highlighted the intrusive nature of this approach to moderating content~\cite{abelson2021bugs}, something our research supports. While the system tested did not limit people from sending messages, we find concern over the ``nannying'' nature of this approach, and how it could invoke a chilling effect~\cite{rosenzweig2020law} whereby people curtail their views, opinions, and beliefs in anticipation of the moderation prompts. This intersects with the design of the system, and in particular the ~\textit{timing}, as during typing interventions evaluate messages before the sender commits to sending. One may equate the draft form of messages as `thoughts', which are now being evaluated by an AI system. Participants spoke of their use of `phantom messages' (i.e., draft messages) to release or vent frustration and anger. 

This highlights the tensions within these systems. On one side, they are designed to change behaviours to prevent harm, and on the other, they may act to silence legitimate discourse. It is important to note that not all participants held these concerns. While most recognised the potential issues related to free speech issues, some felt personally unconcerned; several participants felt strongly towards the free speech and surveillance issues but felt exceptions would be needed when it came to protecting children online, a finding that supports prior research that highlights the effect of crime type on public perceptions of CSS~\cite{geierhaas2023attitudes}. 

\subsection{More than systems of moderation} %7.2
Proactive moderation systems appear to go beyond the simple moderation of content by providing users with education and support tools to help them engage in more effective communications, and identify and learn about changes in language and societal behaviours.  
\subsubsection{Mediating communications}
These tools can act to support people's communications by allowing AI embedded within a moderation system to mediate their communications, in what Hancock et al.~\cite{hancock2020ai} describe as AI-mediated communications (AI-MC). They highlight the potential impact AI-MC can have on interpersonal dynamics, and more specifically on self-presentation, impression formation and trust. Our work supports this, with participants highlighting how the system may impact both how they construct and perceive messages that are mediated through the moderation tool. Some of the effects on interpersonal dynamics are intended outcomes of the system and are aligned with its goals. For example, it aims to impact how messages are constructed to reduce toxicity. Yet, other potential effects are unintended and could have negative implications on users' communications and interpersonal dynamics; for instance, how the knowledge of the system may be used to purposefully signal~\cite{donath2007signals} toxicity intentionality, and how the knowledge of the system may reduce receivers trust in the authenticity of the sender's communications. 

\subsubsection{Educational tools}
We also find these systems have educational value. As language and social behavioural norms change, people highlight the potential for using language or phrases that are toxic, without the intention to do so. While there may be incidents where these tools could educate people, it is also important to highlight prior work that describes how people use covert language such as `dog whistling' to spread hate and avoid moderation~\cite{bhat2020covert}  may plead ignorance around the use of certain terms, claiming the use of toxic language is `accidental'~\cite{muller2007accidental}. Proactive moderation increases the intentionality of comments, reducing the risk of being able to mask these types of behaviours as unintended, as well as educating users who are genuinely unaware of the impact of their use of language, a group that [anonymised for review] describes as the `the unaware' [[hidden for review]].

\subsection{Implications for proactive moderation systems}
We highlight several implications from our findings, whilst highlighting the need for further work around this form of content moderation which we discuss in section~\ref{limitations}. Firstly, as we found no effect of model presentation on moderation efficacy, we suggest designers do not present abstracted outputs from models to avoid negative gamification of the system and to avoid other risks of abuse found in prior work [hidden for review].
%consider presenting abstracted outputs from models to avoid risk of abuse. 
Secondly, as interventions that occur during typing may be more effective where friction is low, where this approach is taken we suggest that models should be refined to reduce the risk of false positives occurring from partially complete messages. However, consideration and further research are needed into the impact typing interventions have on perceptions of surveillance, and the practice of `phantom messaging'. Thirdly, we suggest that these tools can be considered as more than just moderation systems, as they have the potential to support and educate users. As such, we suggest designers consider how they can incorporate educational and communication support elements into these systems, which may result in longer-term behaviour change in users as opposed to short-term changes due to behavioural ``nudges"~\cite{thaler2009nudge,wang2014field}. Finally, our work suggests that these tools could be effective across platforms, in both public and private communication channels. However, designers need to consider the nature of the communication platform, and how moderation prompts may impact communications within specific environments. For example, instant messaging encourages faster communications, while public platforms like Twitter encourage slower less real-time communications. Moderation system design may be impacted by these platform characteristics. 

\section{Limitations and future work} \label{limitations}
While our findings aim to contribute towards a greater understanding of the use of proactive moderation tools across platforms, we highlight several limitations to our work that should be considered when evaluating our findings, and when considering opportunities for future research. Firstly, our work was conducted in a controlled environment which limits its ecological validity. This type of research is difficult without access to platforms, however, with our approach of embedding a system into a keyboard, future researchers could explore the development of a bespoke keyboard to conduct ``in-the-wild'' research. Yet, privacy would need to be considered where AI systems are used. Secondly, longer-term deployments would allow for insights into why and when moderation systems are triggered, their effect across different platforms and types of interactions, and their effect over time. This could help in developing more contextualised models and moderation interactions based on the type of interaction taking place. Thirdly, our experiment resulted in fewer toxic responses from Twitter stimuli when compared to WhatsApp messages. While this may indicate a high level of toxic messages occurring in private platforms, it may also be a result of the stimuli used and so further work is suggested to understand differences between these types of platforms. Our keyboard was limited to text input, with no option to send rich content such as emojis and gifs, which are used regularly within communications~\cite{wiseman2018repurposing,jiang2018perfect}, with memes being used to spread hate~\cite{zannettou2018origins}. Future work should look to include richer communications within moderation designs.

\section{Conclusion}
%Our work contributes to the growing body of support for the proactive content moderation paradigm, support by ever advancing language models for detecting different forms of problematic behaviour. We extend prior work by evaluating different design factors, and exploring these designs with users to understand how they effect communication across different platforms. As a result of our findings, we suggest this moderation approach can do more than moderation, but act as a communication support mechanisms, and help to educate people as language and social behavioural norms change. 

Our work contributes to the growing body of work that supports the proactive content moderation paradigm, aided by advancing language models for detecting different forms of problematic content. Overall, we find proactive moderation prompts significantly reduce message toxicity levels, as post-moderated messages are less toxic than pre-moderated messages.
The interaction of timing and friction could have a significant effect on reducing toxicity. On one hand, different levels of friction (low vs high) can reduce the toxic level of the on-send prompt. On the other hand, intervention timing (typing vs send) has a significant effect when the friction is low. Moreover, people have a positive attitude toward the proactive moderation system, as it improves the safety of online spaces, creates moments for reflection or conscious decision-making before sending messages, better manages senders’ impressions, increases social awareness, and aids in content moderation and restriction. However, they also have concerns about the accuracy of the system, the usability of the system, and communication efficiency. Some controls are required to reduce intrusiveness and prevent privacy issues. These findings provide valuable insights for us to acknowledge our limitations and guide future work. Lastly, our work points to this moderation approach as being more than a system of moderation, as it can also act as a communication support mechanism and help educate people as language and social behavioural norms change.

\begin{acks}
This research was supported by UKRI through REPHRAIN (EP/V011189/1), the UK’s Research Centre on Privacy, Harm Reduction and Adversarial Influence Online.
\end{acks}

%%
%% The next two lines define the bibliography style to be used, and
%% the bibliography file.
\bibliographystyle{ACM-Reference-Format}
\bibliography{references}

%%
%% If your work has an appendix, this is the place to put it.

\end{document}